\documentclass[conference]{IEEEtran}

\IEEEoverridecommandlockouts

\usepackage{cite}
\usepackage{amsmath,amssymb,amsfonts}
\usepackage{algorithmic}
\usepackage{graphicx}
\usepackage{textcomp}
\usepackage{bm}
\usepackage[font=small]{caption}
\usepackage{subfig}
\usepackage{graphicx}
\usepackage{graphics}
\usepackage{epstopdf}
\usepackage{xcolor}
\usepackage{dblfloatfix}    % To enable figures at the bottom of page
\usepackage{verbatim}

\usepackage{subfig}
\usepackage{tabulary}
\usepackage{lipsum}
\usepackage{hyperref}
\usepackage{enumitem}
\usepackage{multicol}
\usepackage{gensymb}

\def\BibTeX{{\rm B\kern-.05em{\sc i\kern-.025em b}\kern-.08em
		T\kern-.1667em\lower.7ex\hbox{E}\kern-.125emX}}

\makeatletter
\newcommand*{\rom}[1]{\expandafter\@slowromancap\romannumeral #1@}
\makeatother
\DeclareMathOperator{\Tr}{Tr}
\renewcommand{\theenumi}{label=\Alph{enumi}}
\begin{document}

\title{Performance of Massive MIMO Self-Backhauling for Ultra-Dense Small Cell Deployments}
\author{\IEEEauthorblockN{Andrea Bonfante\IEEEauthorrefmark{1}\IEEEauthorrefmark{2},
		Lorenzo Galati Giordano\IEEEauthorrefmark{1}, David L\'{o}pez-P\'{e}rez\IEEEauthorrefmark{1},
		Adrian Garcia-Rodriguez\IEEEauthorrefmark{1}, Giovanni Geraci\IEEEauthorrefmark{1}, \\ Paolo Baracca\IEEEauthorrefmark{4}, M. Majid Butt\IEEEauthorrefmark{3}, Merim Dzaferagic\IEEEauthorrefmark{2}, and Nicola Marchetti\IEEEauthorrefmark{2}}\vspace{0.2cm}
\normalsize\IEEEauthorblockA{\IEEEauthorrefmark{1}Nokia Bell Labs Ireland and \IEEEauthorrefmark{4}Nokia Bell Labs Germany}
\IEEEauthorblockA{\IEEEauthorrefmark{2}CONNECT Centre, Trinity College Dublin, Ireland}
\IEEEauthorblockA{\IEEEauthorrefmark{3}University of Glasgow, United Kingdom}
%\normalsize\IEEEauthorblockA{\IEEEauthorrefmark{1}Nokia Bell Labs Ireland, \IEEEauthorrefmark{2}CONNECT Centre, Trinity College Dublin, \IEEEauthorrefmark{4}Nokia Bell Labs Stuttgart, \normalsize\IEEEauthorrefmark{3}University of Glasgow}
\thanks{A. Bonfante was funded by the Irish Research Council and Nokia Ireland Ltd under the grant EPSPG/2016/106. This publication has emanated from research supported in part by a grant from Science Foundation Ireland (SFI) and is co-funded under the European Regional Development Fund under Grant Number 13/RC/2077.}
}

\maketitle
\IEEEpeerreviewmaketitle

\begin{abstract}
	A key aspect of the fifth-generation wireless communication network will be the integration of different services and technologies to provide seamless connectivity. In this paper, we consider using massive multiple-input multiple-output (mMIMO) to provide backhaul links to a dense deployment of \emph{self-backhauling} (s-BH) small cells (SCs) that provide cellular access within the same spectrum resources of the backhaul. Through a comprehensive system-level simulation study, we evaluate the interplay between access and backhaul and the resulting end-to-end user rates. Moreover, we analyze the impact of different SCs deployment strategies, while varying the time resource allocation between radio access and backhaul links. We finally compare the above mMIMO-based \mbox{s-BH} approach to a mMIMO direct access (DA) architecture accounting for the effects of pilot reuse schemes, together with their associated overhead and contamination mitigation effects. The results show that dense SCs deployments supported by mMIMO s-BH provide significant rate improvements for cell-edge users (UEs) in ultra-dense deployments with respect to mMIMO DA, while the latter outperforms mMIMO s-BH from the median UEs' standpoint.
\end{abstract}
	

\section{Introduction} \label{Sec:1}

\begin{figure*}[!t]
	\centering
	\subfloat[Random deployment]{\includegraphics[width=0.9\columnwidth]{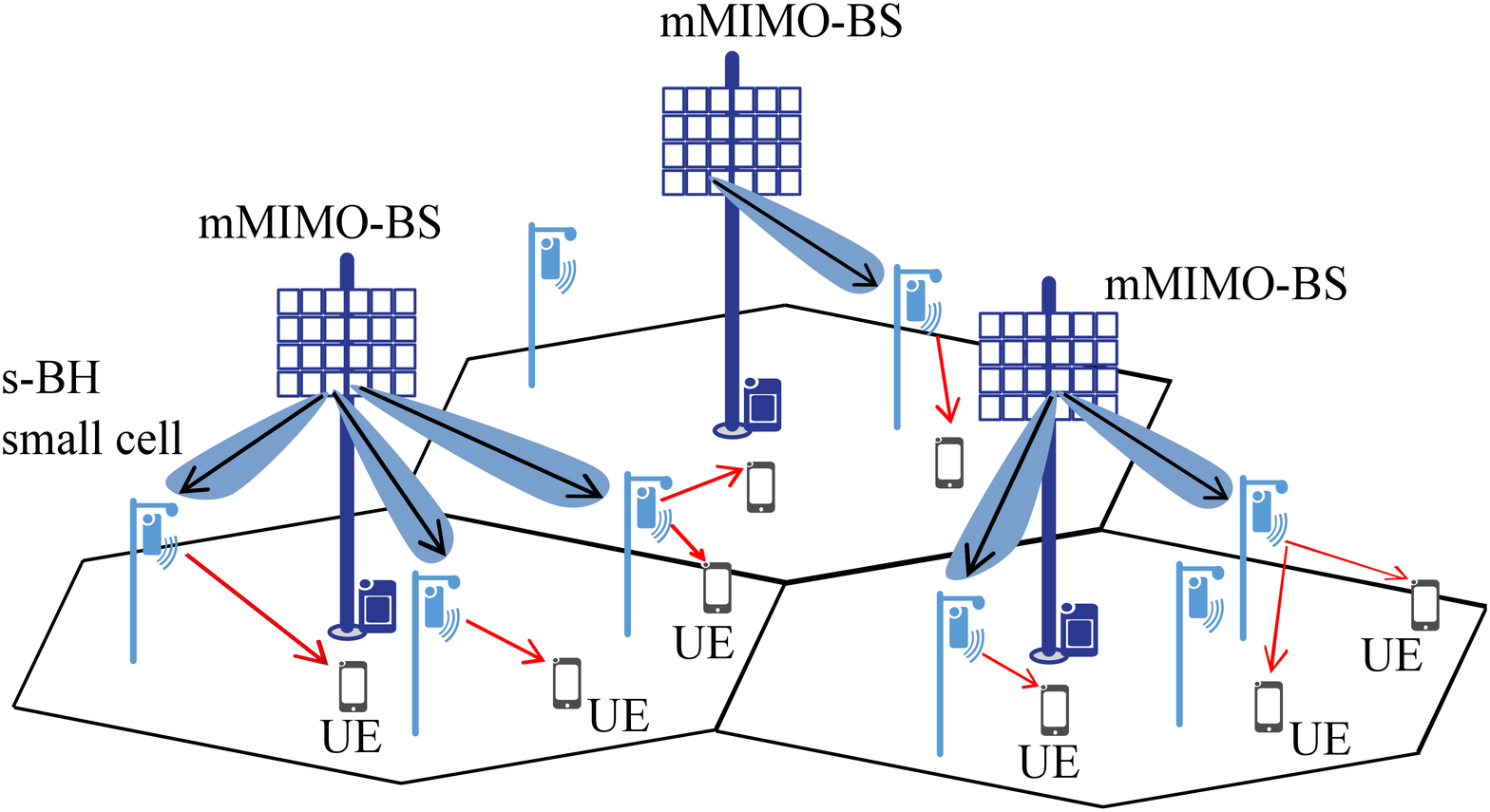}\label{fig2a}}
	\hfill
	\hspace{0.2in}
	\subfloat[Ad-hoc deployment]{\includegraphics[width=0.9\columnwidth]{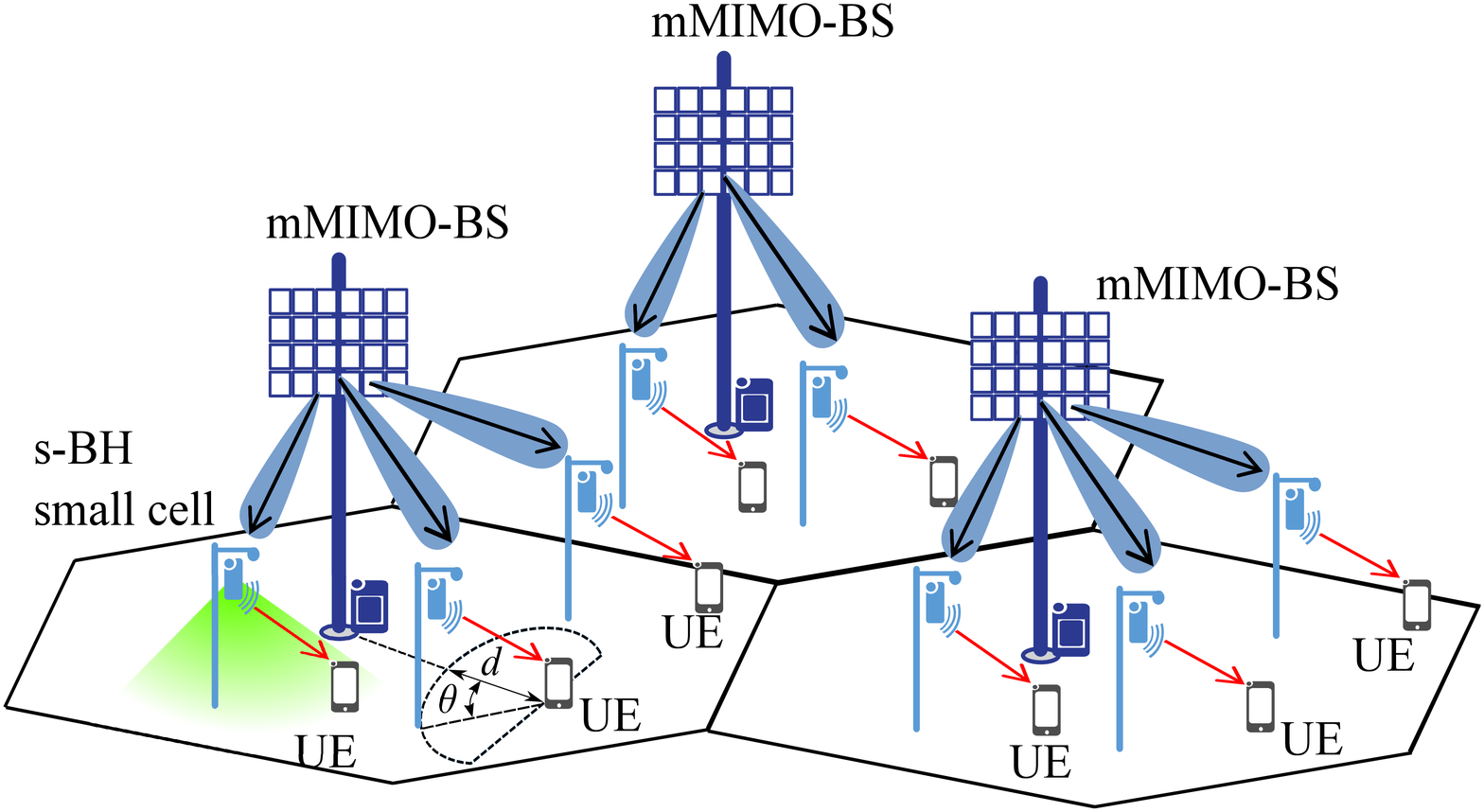}\label{fig2b}}	 
	\caption{Examples of two different SCs deployments considered in the paper.}
	\label{fig:network_layout}
	%\vspace*{-0.3cm}
\end{figure*} 

Fifth-generation (5G) wireless communication systems are expected to support a 1000x increase in capacity compared to existing networks \cite{7169508}. Meeting this gargantuan target will require mobile network operators (MNOs) to leverage new technologies such as massive multiple-input multiple-output (mMIMO), and deploy a large number of additional small cell base stations \cite{6375940, 7126919}. Wireless self-backhauling (\mbox{s-BH}), achieved through the tight integration of these two complementary means, lures MNOs with the potential of achieving the desired capacity boost at a contained investment \cite{7306534}. Indeed, exploiting the large number of spatial degrees-of-freedom available with mMIMO to provide sub-6\,GHz in-band wireless backhauling to small cells (SCs) offers multiple advantages to MNOs: avoiding deployment of an expensive wired backhaul infrastructure, availing of more flexibility in the deployment of SCs, and not having to purchase additional licensed spectrum as in the case of out-of-band wireless backhauling \cite{7306536}.

The integration of radio access and backhauling -- advocated in s-BH solutions -- has been addressed by the Third Generation Partnership Project (3GPP) with a list of requirements detailed in \cite{3gpp.TS-22.261}. At the same time, various research efforts have tackled the problem of resource allocation and management in s-BH networks in the time and frequency domains \cite{Gupta,Nguyen2016}. Finally, the combination of mMIMO spatial multiplexing and s-BH has been considered in \cite{Kela2017,7445888_GioBackhaul,8241817} and also in \cite{Tabassum,7177124,7996532}, albeit for full-duplex scenarios.

In this paper, we analyze the end-to-end user equipment (UE) performance of mMIMO s-BH networks. In particular, we consider a realistic multi-cell setup where mMIMO base stations (mMIMO-BSs) provide sub-6\,GHz backhauling to a plurality of half-duplex SCs overlaying the macro cellular area. We evaluate the UE data rates achieved through s-BH in two ultra-dense deployment scenarios, namely a \emph{random} deployment -- where SCs are uniformly distributed over a geographical area --, and an \emph{ad-hoc} deployment -- where SCs are purposely positioned close to UEs to achieve line-of-sight (LOS) access links. In these half-duplex systems, a s-BH approach entails sharing time-and-frequency resources between radio access and backhaul links. To the best of the authors’ knowledge,
in this paper, we also compare for the first time the performance of the mMIMO s-BH approach and that of a direct access (DA) approach, where mMIMO-BSs are solely dedicated to serving UEs in the absence of SCs \cite{Lim}.
Our study provides a number of key takeaways:
\begin{itemize}
	\item Adding more \emph{randomly} deployed SCs -- where SCs are randomly placed -- provides limited gains for the end-to-end UE rates. The UEs benefit from more radio resources allocated and from the SCs proximity, given by the higher probability of the UEs to be close to -- and in LOS with -- the serving SCs. However, the UEs are affected by a significantly higher inter-cell interference because they see a growing number of interfering links in LOS conditions. Overall, the detrimental impact of interference outweighs the combination of the gains provided by more radio resources and proximity, preventing the UEs to take the advantages of the SCs dense deployment.
	%\item Adding more \emph{randomly} deployed SCs -- where SCs are randomly placed -- provides limited gains for the end-to-end UE rates, because of high inter-cell interference. Indeed, although UEs experience a higher probability of being close to -- and in LOS with -- the serving SC, they also see a growing number of interfering links in LOS conditions. Nevertheless, a denser deployment implies that each UE can have allocated more physical resource blocks (RBs) on the access link, as the number of UEs connect to each SC tends to one. 
	
	\item Adding \emph{ad-hoc} deployed SCs -- where SCs are placed in proximity to UEs -- provides higher data rates, thanks to a high signal-to-interference-plus-noise ratio (SINR) on the access link, given by the higher proximity gains with respect to the random deployment. %This deployment gives an upper-bound for the end-to-end UE rates, to which converge with an ultra-dense deployment of SCs placed randomly.
	
	\item Partitioning resources between wireless access and backhaul links is of paramount importance. Indeed, the end-to-end performance is sensitive to said partition, and optimal rates can only be achieved through a carefully designed tradeoff.
	
	\item Unlike mMIMO s-BH -- where mMIMO-to-SC links are static, and thus channel acquisition is facilitated -- mMIMO DA suffers more from pilot overhead and contamination. Indeed, when compared to mMIMO DA solutions with pilot reuse 3 and reuse 1, ultra-dense SCs deployments supported by mMIMO s-BH provide rate improvements for cell-edge UEs that amount to $30\%$ and a tenfold gain, respectively. On the other hand, mMIMO DA outperforms s-BH from the median UEs' standpoint.
\end{itemize}

\textit{Notation:} Capital and lower-case bold letters denote matrices and vectors, respectively, while $[\cdot]^*$, $[\cdot]^\mathrm{T}$ and $[\cdot]^\mathrm{H}$ denote conjugate, transpose, and conjugate transpose, respectively.

\section{System Model} \label{Sec:2}

As shown in Fig. \ref{fig:network_layout}, we focus on the study of the downlink (DL) performance for a two-tier heterogeneous network formed by mMIMO-BSs overlaying a layer of self-backhauled SCs. The mMIMO-BSs are connected to the core network through a high-capacity wired connection, while all SCs receive backhaul traffic through mMIMO-BSs and function as access points for UEs. We consider a self-backhaul configuration, where mMIMO-BSs are solely dedicated for the backhaul, while SCs are solely dedicated for the access. For comparison purposes, we also consider the conventional DA approach where each mMIMO-BS directly serves the UEs.

\subsection{Macro cell and user topologies}
We denote by $\mathcal{I}$ the set of mMIMO-BSs placed in a uniform hexagonal grid with three sectors per site. Each mMIMO-BS $i$, is equipped with a large number of antennas $M$, and serves $L_{i}$ single-antenna SCs. Furthermore, we denote by $K_{i}$ the number of UEs randomly and uniformly distributed over the sector's area, and let $k$ denotes single-antenna UE. We assume that each UE is connected with the SC (in the s-BH approach) or with the mMIMO-BS (in the DA approach) that provides the largest reference signal received power (RSRP) \cite{3gpp.36.814}.

\subsection{Small cell topologies}\label{subs:sc-dep}
We denote by $\mathcal{L}_{i}$ the set of SCs deployed per sector and connected to the $i$-th mMIMO-BS that provides the largest RSRP. Each SC connects $K_{l}$ UEs. Two different SCs deployments are presented in the following:
%$L_{i}$ denote the number of SCs in the set $\mathcal{L}_{i}$

\begin{enumerate}[label={(\alph*)}]
	\item \textbf{Random deployment:} Self-backhauled SCs are randomly and uniformly distributed over the mMIMO-BS geographical area as shown in Fig. \ref{fig2a}. This scenario is used as a baseline and follows the set of parameters specified by the 3GPP in \cite{3gpp.36.814} to evaluate the relay scenario.
	
	\item \textbf{Ad-hoc deployment:} Self-backhauled SCs are positioned targeting nearby UE locations. This scenario is used as an example of ultra-dense network deployment. We assume the possibility to realize this target of network deployment, for example by means of drone-BSs, where the drone-BSs can reposition following the locations of UEs \cite{7744808}.\footnote{Although mentioned, the drone-BSs use-case is not the focus of this paper and it is left for future investigation, since the height of the outdoor SC antenna is fixed to 5 meters, and we use the channel models adopted for the relay study \cite{3gpp.36.814}.}	
	%Moreover, this represents a realistic scenario to connect static and low-mobility UEs, whereas medium and high mobility UEs can be directly connected to the mMIMO-BS \cite{7070656}.
	As shown in Fig. \ref{fig2b}, we model this scenario by considering SCs deployed within a 2-D (two-dimensional) distance $d$ of the UEs, and an angle $\theta$ measured from the straight segment that links UEs and their closest mMIMO-BS. $\theta$ is chosen uniformly at random from $-\pi/2$ and $\pi/2$. It is worth noting that even when the 2-D distance $d=0$, UEs and SCs are still separated in space because the antennas are positioned at different heights. More precisely, they are assumed located at 1.5 meters and 5 meters above the ground, for the UEs and the SCs, respectively \cite{3gpp.36.814}. \label{ad-hoc}
\end{enumerate}

With a dense deployment of SCs, the UE SINRs are severely affected by the strong inter-cell interference among SCs. 
In addition, to limit the effect of the inter-cell interference, with the ad-hoc deployment, we propose to replace at the SC the isotropic antenna (Patch antenna) with a more directive antenna (Yagi antenna) pointing downwards to the ground (as shown by the green radiation cone in Fig. \ref{fig2b}), and therefore only illuminating the closest UEs: details about this modeling can be found in Table \ref{table:parameters}.

\subsection{Frame structure} \label{SubSec: frame}

\begin{figure}[!t]
	\centering
	\subfloat[]{\includegraphics[width=\linewidth]{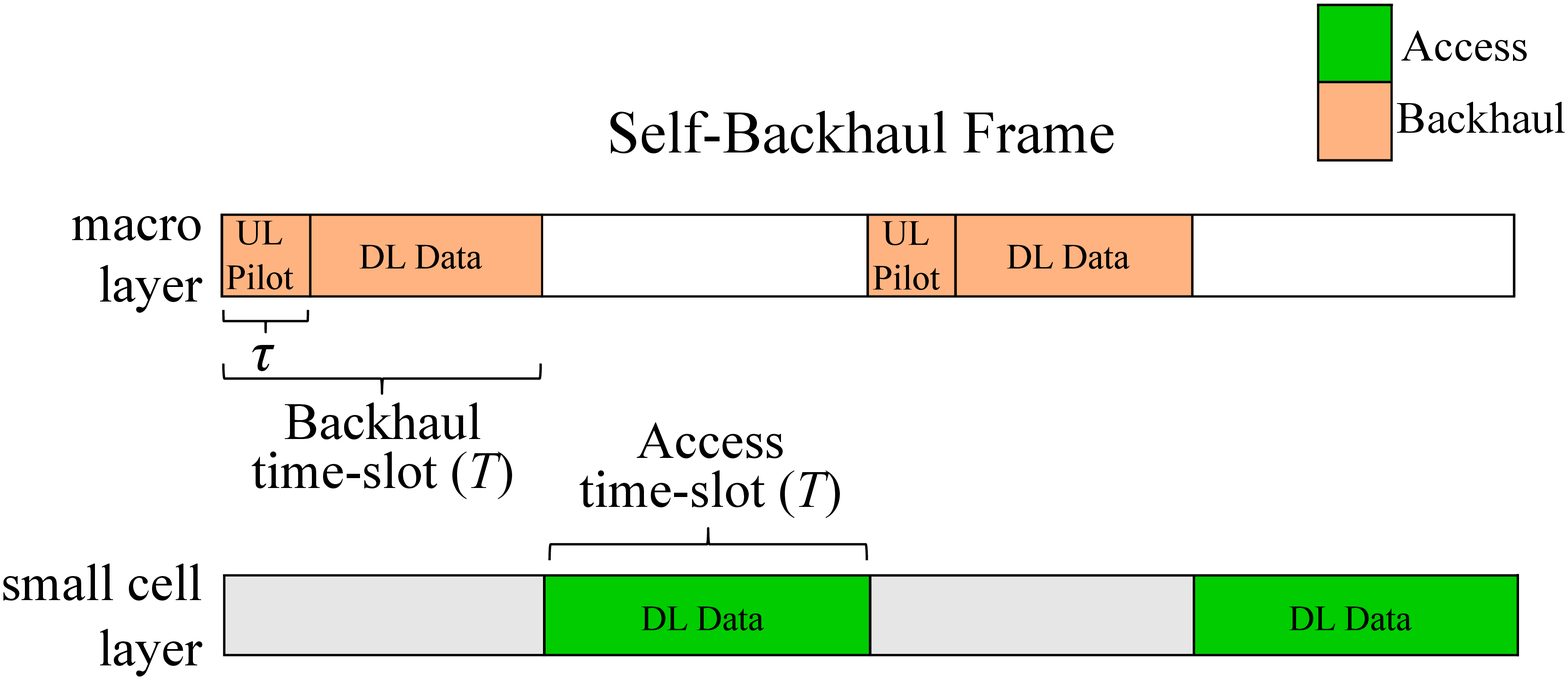}\label{fig:3a}}\hfill
	\subfloat[]{\includegraphics[width=\linewidth]{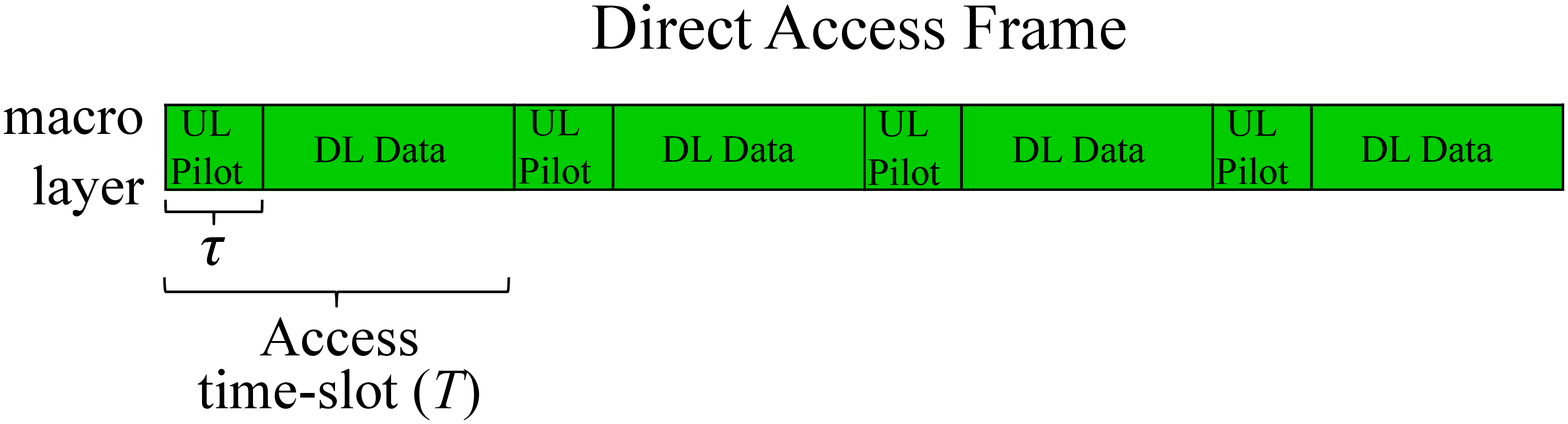}\label{fig:3b}}
	\caption{DL frame structure for mMIMO s-BH with $\alpha = 0.5$ (Fig. \ref{fig:3a}) and for mMIMO DA (Fig. \ref{fig:3b}).}
	\vspace*{-0.2cm}
	\label{fig:frame structure}
\end{figure}

As shown in Fig. \ref{fig:3a}, we consider the time-slot $T$ as a single scheduling unit in the time domain, and we partition the access and backhauling resources through the parameter $\alpha \in [0,1]$. Therefore, $\alpha$ time-slots are allocated to the backhaul links, while $1-\alpha$ time-slots are allocated to the access links. In the frequency domain, we divide the system bandwidth $B$ into $Q_{t}$ resource blocks (RBs), and we allocate all the RBs to the backhaul links or the access links. We make the following assumptions in considering the partition of backhaul and access time-slots among the SCs and UEs:

\begin{itemize}[noitemsep,topsep=0pt]
	\item During the backhaul time-slots, all the associated SCs are served by the mMIMO-BS $i$, and we use the same value of $\alpha$ for all the SCs. 
	In this approach, the mMIMO-BSs precode the backhaul signals towards the single-antenna SCs, which are spatially multiplexed in the same time-frequency resources by allocating all the RBs in $T$ to each SC. 
	\item During the access time-slots, the SCs schedule their connected UEs by using a Round Robin (RR) mechanism as frequency domain scheduler. This approach entails that each SC equally shares the system bandwidth $B$ with its UEs. 
\end{itemize}

In the Fig. \ref{fig:3b}, it is shown the frame structure used for the DA setup, where all the time-slots are allocated to the access links. In each time slot, the mMIMO BSs precode the access signals, and the UEs are spatially multiplexed on the entire system bandwidth.
Figs. \ref{fig:3a} and \ref{fig:3b} also show the fraction $\tau$ of the time-slots dedicated for the transmission of the pilot sequences, used to estimate the massive MIMO channel. Details about the channel training procedure will be discussed in Section \ref{Sec:3}.

\subsection{Channel model}

We define as $\bm{\mathrm{h}}_{il}=[h_{il1},\ldots,h_{ilM}]^{\mathrm{T}} \in \mathbb{C}^{M}$ the propagation channel between the $l$-th single-antenna receiver (SC in the s-BH architecture and UE in the mMIMO DA) and the $M$ antennas of the $i$-th mMIMO-BS. The composite channel matrix between the $i$-th mMIMO-BS and the devices in the $i'$-th cell is represented by $\bm{\mathrm{{H}}}_{i, i'} = [\bm{\mathrm{{h}}}_{i1} \cdots \bm{\mathrm{{h}}}_{iL_{i'}}] \in \mathbb{C}^{M \times L_{i'}} $. Since all the RBs are assigned to each SC, we removed the RB index $q$ from the massive MIMO channel notation.

Furthermore, we define as $g_{lkq} \in \mathbb{C} $ the single-input single-output (SISO) channel between the $l$-th SC and the $k$-th UE in the $q$-th RB. Each channel coefficient $h_{ilm} = \sqrt{ \beta_{il}} \tilde{h}_{ilm}$, and $g_{lkq} = \sqrt{ \beta_{lk}} \tilde{g}_{lkq}$ accounts for both the effects of a large scale fading and a small scale fading components: 
\begin{itemize}[noitemsep,topsep=0pt]
	\item The large fading components $\beta_{il},\beta_{lk} \in \mathbb{R}^{+}$ have been modeled by using a combined LOS/Non-LOS (NLOS) path loss model, which accounts for the shadowing effect, set to be log-normal distributed with different standard deviations \cite{3gpp.36.814}. Because of its slow-varying characteristic, it does not change rapidly with time, and it can be assumed constant over the observation time-scale of the network.
	\item The small scale fading components $\tilde{h}_{ilm}, \tilde{g}_{lkq} \in \mathbb{C}$, which results from multi-path, have been modeled as a Rician fast-fading, which rapidly changes over time and frequency. For the LOS channels, we characterize the Rician $K$ factor with the model: $K [dB] = 13-0.03r$  in dB, where $r$ is the distance between transmitter and receiver in meters \cite{3gpp.25.996}.
\end{itemize}
Throughout the paper, we assume a composite fading (i.e. large scale fading and small scale fading together) for the SC-UE and the mMIMO-BS-UE links (in the DA approach), which changes between successive time-slots and between different RBs. Moreover, because of the static position of the SCs, we consider that the backhaul channel SC-mMIMO-BS remains constant for a period $T_{BH} \gg T$.

\section{End-to-end UE rates} 
\label{Sec:3}

In this section, we provide the detailed description of the operations required for the DL transmission in the mMIMO s-BH approach and in the mMIMO DA approach. We describe the channel training procedure, the mMIMO DL backhaul transmission, and the DL access transmission, which is treated separately below for both the s-BH and the mMIMO DA setups.

\subsection{Massive MIMO channel training}

To calculate the DL precoder, we consider that the channel is estimated through uplink (UL) pilots, assuming UL/DL channel reciprocity \cite{6375940}. We also assume that the SCs or UEs associated to the same mMIMO-BS have orthogonal pilot sequences, and define the pilot code-book with the matrix $\bm{\mathrm{\Phi}}_{i} = [ \bm{\mathrm{\phi}}_{i1} \cdots \bm{\mathrm{\phi}}_{iL_{i}} ]^{\mathrm{T}}  \in \mathbb{C}^{L_{i} \times  S}$ , which satisfies $ \bm{\mathrm{\Phi}}_{i} \bm{\mathrm{\Phi}}_{i}^\mathrm{H} = \bm{\mathrm{I}}_{L_{i}}$. Here, the $l$-th sequence is given by $\bm{\mathrm{\phi}}_{il} = [\phi_{il1},\ldots,\phi_{ilS}]^{\mathrm{T}} \in \mathbb{C}^{S}$, and $S$ denotes the pilot code-book length. Note that $L_{i} \leq S$, i.e., the maximum number of SCs or UEs served by the mMIMO-BSs in a time-slot is limited by the number of orthogonal pilot sequences.
The matrix $\bm{\mathrm{Y}}_{i} \in \mathbb{C}^{M \times  S}$ of pilot sequences received at the $i$-th mMIMO-BS can be expressed as \cite{Zhu2016}
\begin{equation} \label{eq: ul}
\bm{\mathrm{Y}}_{i} = \sqrt{P_{il}^{\mathrm{ul}}} \sum\limits_{i' \in \mathcal{I}} \bm{\mathrm{{H}}}_{i,i'} \bm{\mathrm{\Phi}}_{i'} + \bm{\mathrm{{N}}}_{i},
\end{equation}
where $P_{il}^{\mathrm{ul}}$ is the power used by the $l$-th device located in the $i$-th sector for UL pilot transmission, and $\bm{\mathrm{{N}}}_{i} \in \mathbb{C}^{M \times  S}$ represents an additive noise, and is modeled with independent and identically distributed complex Gaussian random variable.

Let $\bm{\mathrm{{H}}}_{i}$ denote the channel between the $i$-th mMIMO-BS and the UEs located in the same sector. During the UL training phase, the mMIMO-BS obtains an estimate of $\bm{\mathrm{{H}}}_{i}$ by correlating the received signal with a known pilot matrix $\bm{\mathrm{\Phi}}_{i}$. Let us define $\mathcal{P} \subseteq \mathcal{I}$ as the subset of sectors, whose UEs share identical pilot sequences with the UEs served by the $i$-th mMIMO-BS. The resulting estimated channel can be expressed as
\begin{equation} \label{eq: ch est}
\bm{\mathrm{\widehat{H}}}_{i} = \frac{1}{\sqrt{P_{il}^{\mathrm{ul}}}} \bm{\mathrm{Y}}_{i}  \bm{\mathrm{\Phi}}_{i}^{\mathrm{H}} = \bm{\mathrm{{H}}}_{i} + \sum_{i' \in \mathcal{P}} \bm{\mathrm{{H}}}_{i,i'} + \frac{1}{\sqrt{P_{il}^{\mathrm{ul}}}} \bm{\mathrm{{N}}}_{i} \bm{\mathrm{\Phi}}_{i}^{\mathrm{H}}.
\end{equation}
The first, second and third terms on the right hand side of \eqref{eq: ch est} represent the estimated channel, a residual pilot contamination component and the noise after the correlation, respectively. The use of the same set of orthogonal pilot sequences among different sectors leads to the well-known pilot contamination problem, which can severely degrade the performance of mMIMO systems \cite{6375940, Galati1712}. 
In this paper, we assume that no pilot contamination occurs for the mMIMO s-BH system. 
%This happens because the set of orthogonal pilot sequences can span multiple time-slots due to the longer coherence time of the static backhaul channel, $T_{BH}$, with respect to the access time-slot, $T$, therefore avoiding pilot reuse. 
Due to the longer coherence time of the static backhaul channel, $T_{BH}$, with respect to the system time-slot, $T$, mMIMO pilots do not need to be transmitted in every time-slot dedicated to backhauling, thus allowing higher reuse factors with fully orthogonality over the entire network. 
In contrast, for mMIMO DA this assumption does not hold and, in this paper, we consider that a maximum of 16 orthogonal pilot sequences can be multiplexed in a single orthogonal frequency division multiplexing (OFDM) symbol \cite{Galati1712}. 
In both mMIMO s-BH and mMIMO DA, the overhead associated to the UL training phase are considered and measured in terms of number of OFDM symbols $\tau$.
%the number of OFDM symbols allocated to the UL training, 
Two pilot allocation schemes are here compared:

\begin{itemize}
	\item \textit{Pilot reuse 1 scheme (R1):} All $K_{i}$ UEs per sector are trained in $\tau=1$ OFDM symbol.
	
	\item \textit{Pilot reuse 3 scheme (R3):} The sectors of the same site use orthogonal pilot sequences. This scheme avoids pilot contamination from co-sited sectors, but requires $\tau=3$ OFDM symbols, resulting in a higher pilot overhead when compared to the R1 scheme.
\end{itemize} 

\subsection{Massive MIMO s-BH DL transmission}

The $i$-th mMIMO-BS uses the precoding matrix $\bm{\mathrm{W}}_{i}=[\bm{\mathrm{w}}_{i1} \cdots \bm{\mathrm{{w}}}_{iL_{i}}] \in \mathbb{C}^{M \times L_{i}}$ to serve its connected UEs during the DL data transmission phase. In this paper, we consider that $\bm{\mathrm{W}}_{i}$ is computed based on the zero-forcing (ZF) criterion as
\begin{equation} \label{eq: zf}
\bm{\mathrm{W}}_{i} = {\bm{\mathrm{D}}_{i}}^{\frac{1}{2}} \bm{\mathrm{\widehat{H}}}_{i} \left({\bm{\mathrm{\widehat{H}}}^{\mathrm{H}}_{i}} {\bm{\mathrm{\widehat{H}}}_{i}} \right) ^{-1}.
\end{equation} 
Here, the diagonal matrix $\bm{\mathrm{D}}_{i} = \operatorname{diag} \left(\rho_{i1},\rho_{i2},\ldots,\rho_{iL_{i}} \right)$ is chosen to equally distribute the total DL power $P_{i}^{\mathrm{dl}}$ among the $L_{i}$ receivers. In the previous expression, $\rho_{il}$ represents the power allocated to the $l$-th receiver located in the $i$-th sector, and $\Tr\{\bm{\mathrm{D}}_{i} \}=P_{i}^{\mathrm{dl}}$, where $\Tr\{\bm{\mathrm{D}}_{i} \}$ is the trace of matrix $\bm{\mathrm{D}}_{i}$.
%The above entails that $P_{i}^{\mathrm{dl}} = \sum_{l=1}^{L_{i}}{ \rVert \bm{\mathrm{w}}_{il} \rVert}^2$, where ${ \rVert \bm{\mathrm{w}}_{il} \rVert}$ indicates the norm of $\bm{\mathrm{w}}_{il}$. %TOREMOVE

The SINR of the $l$-th DL stream transmitted by the $i$-th mMIMO-BS can be expressed as
\begin{equation} \label{eq: sinr MIMO}
\mathrm{SINR}_{il} = \dfrac{ \rho_{il} | \bm{\mathrm{h}}^{\mathrm{H}}_{il} \bm{\mathrm{w}}_{il} |^2}
{ \sum\limits_{\substack{j \in \mathcal{L}_{i} \\ j\neq l}}
	{ \rho_{ij} | \bm{\mathrm{h}}^{\mathrm{H}}_{il} \bm{\mathrm{w}}_{ij} |^2} + \sum\limits_{\substack{i' \in \mathcal{I}  \\ i' \neq i}} \sum\limits_{j \in \mathcal{L}_{i'}} {\rho_{i'j} | \bm{\mathrm{h}}^{\mathrm{H}}_{i'l} \bm{\mathrm{w}}_{i'j}|^2} + \sigma^2_{n}}.
\end{equation}
The numerator of \eqref{eq: sinr MIMO} contains the power of unit-variance signal intended for the $l$-th receiver, 
while the denominator includes the co-channel interference from the serving $i$-th mMIMO-BS, the inter-cell interference from other mMIMO-BSs, and the power of the thermal noise at the receiver $\sigma^2_{n}$.

The corresponding DL backhauling rate at the $l$-th SC receiver can therefore be expressed as
\begin{equation} \label{eq: rate mMIMO BH}
R_{il}^{\mathrm{BH}} = \left(1 - \frac{\tau}{T} \right) B \log_2 \left( 1 + \mathrm{SINR}_{il} \right).
\end{equation}

\subsection{Small cell DL transmission}

We recall from the channel model that $g_{lkq}$ denotes the SISO channel  between the $l$-th SC and the $k$-th UE corresponding to the $q$-th RB. The SINR of the $k$-th UE served by the $l$-th SC in RB $q$ can be expressed as
\begin{equation} \label{eq: SINR access}
\mathrm{SINR}_{lkq} = \dfrac{ P_{l}^{\mathrm{dl}} |g_{lkq}|^2}
{ \sum\limits_{i \in \mathcal{I}} \sum\limits_{\substack{l' \in \mathcal{L}_{i}  \\ l' \neq l}} P_{l'}^{\mathrm{dl}} |g_{l'kq}|^2  + \sigma^2_{n_{2}}},
\end{equation} 
where $P_{l}^{\mathrm{dl}}$ and $P_{l'}^{\mathrm{dl}}$ are the transmit powers on the RB of the $l$-th and $l'$-th SCs, respectively, and $\sigma^2_{n_{2}}$ denotes the thermal noise power at the UE receiver. 
% We assume a fixed transmit power at each SC.
The DL access rate for UE $k$ served by SC $l$ can be therefore expressed as
\begin{equation} \label{eq: rate access}
R_{lk}^{\mathrm{AC}} = \frac{B}{Q_{t}} \sum_{q=1}^{Q_{t}} x_{q}^{k} \log_2 \left( 1 + \mathrm{SINR}_{lkq} \right),
\end{equation} 
where $x_{q}^{k}= 1$ if the $q$-th resource block is assigned to the $k$-th user, and $x_{q}^{k}= 0$ otherwise. The aggregated DL access rate provided by the $l$-th SC is $R_{l}^{\mathrm{AC}} = \sum_{k=1}^{K_{l}} R_{lk}^{\mathrm{AC}}$. The actual aggregated DL access rate provided by the $l$-th SC depends on the backhaul DL rate, which entails that $R_{l}^{\mathrm{AC}} \leq R_{il}^{\mathrm{BH}}$, $\forall l \in \mathcal{L}_{i}$, and $\forall i \in \mathcal{I}$. 
In this paper, we assume that the backhaul capacity is equally divided between the $K_{l}$ UEs served by the $l$-th SC.\footnote{The assumption of equally distributed backhaul capacity might become a drawback for the end-to-end rates when UEs served by the same SC have significant differences between the rates of the access links, and in this case, the partition of the backhaul resources among the UEs could be designed proportionally to their access rates. This access-based partition of the backhaul resources among the UEs is not the focus of this paper, and its study in the context of self-backhaul is left for future work.} Therefore, the resulting end-to-end access rate for the $k$-th UE can be expressed as
\begin{equation} \label{eq: rate e2e}
R_{ilk} = \min \left( \alpha \frac{R_{il}^{\mathrm{BH}}}{K_{l}}, (1-\alpha) R_{lk}^{\mathrm{AC}}  \right),
\end{equation}
where $\alpha$, as indicated before, represents the time-slots allocated to the backhaul links. 

\subsection{Massive MIMO direct access transmission}

In contrast to s-BH setups, mMIMO systems providing DA dedicate all their time resources to DL data transmission. Therefore, the DL access rate of the $k$-th UE served by the $i$-th mMIMO-BS can be expressed as
\begin{equation} \label{eq: rate mMIMO AC}
R_{ik}^{\mathrm{AC}} = \left(1 - \frac{\tau}{T} \right) B \log_2 \left( 1 + \mathrm{SINR}_{ik} \right),
\end{equation}
where the estimated channel matrix $\bm{\mathrm{\widehat{H}}}_{i} = [\bm{\mathrm{\widehat{h}}}_{i1} \cdots \bm{\mathrm{\widehat{h}}}_{iK_{i}}] \in \mathbb{C}^{M \times K_{i}} $ between the $i$-th mMIMO-BS and its connected UEs is plugged into \eqref{eq: zf}, to subsequently derive \eqref{eq: sinr MIMO} and \eqref{eq: rate mMIMO AC}.

\section{Numerical Results} \label{Sec:4}

To realistically evaluate the mMIMO s-BH network performance, in this paper, we adopt the methodology described by 3GPP in \cite{3gpp.36.814} for heterogeneous network. We perform system level simulations accounting for all signal and interfering radio links between each SC and the UEs, as well as between each mMIMO-BS and all SCs. We collect statistics for different network realizations, each with independent deployments of UEs and SCs. Subsequently, we measure the performance in terms of cumulative distribution function (CDF) of the end-to-end UE rate \eqref {eq: rate e2e}. To compare s-BH against DA, we also simulate the links between mMIMO-BSs and UEs, and compute the resultant rates \eqref{eq: rate mMIMO AC}. Table \ref{table:parameters} contains the relevant parameters used to conduct the simulation campaign.

% [Adrian:] TABLE COMMENTS:
% [Adrian:] It would be good to explicitly capture that there is a LOS probability in the model, even if we need to repeat what is written for the path loss and shadowing.
% [Adrian:] QUESTION. Aren't there more variables that we could add here so that the reviewers understand the accuracy of our simulations? (Scheduling, etc.). We should highlight that the scenario is very complex to simulate!!!!
% [Adrian]: SUGGESTION:  Please add references so that it's clear from where some parameters were extracted (for example, the patch and yaji)
% [Adrian]: QUESTION. Which is $T_{BH}$ in the simulations? This should be included in the table.
%TODO_ab

\begin{table}[!t]
	\vspace{5mm} 
	\centering
	\caption{Simulation parameters}
	\vspace{-2mm} 
	\label{table:parameters}
	\begin{tabulary}{\columnwidth}{ |p{2.9cm}| p{5cm}  | }
		\hline
		\textbf{mMIMO-BSs}		& \textbf{Description} \\ \hline
		Cellular layout				& Wrap around hexagonal, 19 sites, 3~sectors/site \\ \hline
		Deployment				    & Inter-site distance: $500$~m, height: $32$~m   \cite{3gpp.36.814} \\ \hline
		Antenna array 				& Uniform linear, spacing: $0.5\lambda$, Number of antennas per array: 64 \\ \hline
		Antenna pattern 			&  $70^{\circ}$ H x $10^{\circ}$ V beamwidths, 14 $\mathrm{dBi}$ max., downtilt: $15^{\circ}$  \cite{3gpp.36.814} \\ \hline
		Precoder					& Zero-forcing \\ \hline
		Tx power/Noise figure		& 46 dBm, 5 dBm \cite{3gpp.36.814}  \\ \hline 
		\textbf{Self-BH SCs} 		& \textbf{Description} \\ \hline
		Deployment  				& Random: $\{4,8,16\}$ SCs/sector, Ad-hoc: 16 SCs/sector, height: $5$~m \\ \hline
		Backhaul antenna pattern 	& 5 $\mathrm{dBi}$ antenna gain,  Omni \cite{3gpp.36.814} \\ \hline
		Access antenna pattern  -- Patch	&  $80^{\circ}$ H x $80^{\circ}$ V beamwidths, 5 $\mathrm{dBi}$ max., downtilt: $90^{\circ}$ \\ \hline
		Access antenna pattern -- Yagi	& $58^{\circ}$ H x $47^{\circ}$ V beamwidths, 10 $\mathrm{dBi}$ max., downtilt: $90^{\circ}$ \\ \hline
		Tx power/Noise figure		& 30 dBm, 5 dB \cite{3gpp.36.814} \\ \hline
		\textbf{UEs}				& \textbf{Description} \\ \hline
		Deployment 				    & Random, 16 UEs/sector on average, all served, height: $1.5$~m \\ \hline
		Tx power/Noise figure     	& 23 dBm, 9 dB \cite{3gpp.36.814} \\ \hline
		\textbf{Channel} 			& \textbf{Description} \\ \hline
		Scenario 				    & Outdoor SCs, outdoor UEs \\ \hline
		Bandwidth/Time-slot 		& 10 MHz at 2 GHz, $Q_{t} = 50$ RBs, $T=1$ msec. \cite{3gpp.36.814} \\ \hline
		LOS probability, path loss and shadowing	    & mMIMO-BS to UE (based on 3GPP macro to UE models), mMIMO-BS to SC (based on 3GPP macro to relay models), SC to UE (based on 3GPP relay to UE models) as per \cite{3gpp.36.814} \\ \hline
		Fast fading  				& Rician, distance-dependent K-factor \cite{3gpp.25.996} \\ \hline
		Thermal noise 				& -174 dBm/Hz power spectral density \\ \hline
	\end{tabulary}
	\vspace{-5mm} 
\end{table}

\subsection{Small cell random and ad-hoc deployments with mMIMO s-BH} 

\begin{figure}[!t]
	\centering
	\includegraphics[width=0.9\columnwidth]{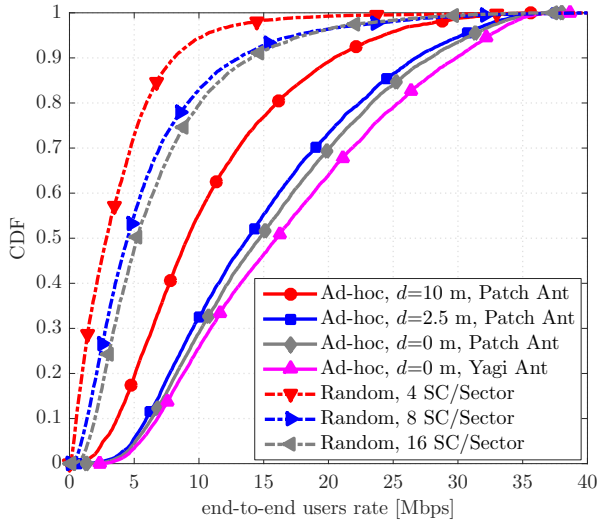}
	\caption{CDF of end-to-end UE rate in: ($i$) ad-hoc deployment of 16 SCs per sector with variable UE-to-SC distance $d$ and different antenna patterns (Patch and Yagi); ($ii$) random deployment of SCs.}
	\label{fig:4}
\end{figure}

In Fig. \ref{fig:4}, we assume $\alpha=0.5$, and analyze the results for the two SC topologies described in Sec. \ref{subs:sc-dep}, namely the ad-hoc and random SC deployments. In both cases, $K_{i}=16$ UEs are deployed per sector, and scheduled in access time-slots by their serving SCs. We evaluate the impact of densification by considering $L_{i}=\{4,8,16\}$ SCs per sector for the case of random SC deployments. In the ad-hoc deployment, we consider $L_{i}=16$ SCs per sector, and different values of the 2-D UE-to-SC distance $d$.

The results of Fig. \ref{fig:4} illustrate that the improvements attained by adding more SCs in the random deployment scenario are limited. This occurs because when densifying the network the carrier signal benefits from having SCs that are more likely in close vicinity with the served UE.
Moreover, the UEs benefit from more radio resources allocated. In fact, with more SCs, there are less UEs served per SC, and the SC can allocate more RBs per UE. 
However, adding more SCs increase the probability of having a larger number of interfering SCs with a LOS channel with respect to the UE.
As a result, the power of the interference links grows faster than the carrier signal power due to NLOS to LOS transition of the interference links \cite{7335646}. 
The gains provided by more radio resources and proximity are outweighed by the detrimental impact of interference, and from the curves shown in Fig. \ref{fig:4}, we can see that the end-to-end UE rates increase marginally when doubling the number of SCs deployed.  

Regarding the ad-hoc deployment, the results of Fig. \ref{fig:4} demonstrate that decreasing $d$ leads to significant improvements in the UEs rates. This is because UEs benefit from a robust LOS component and reduced path loss, which increases signal power gains and consequently leads to larger SINRs. Fig. \ref{fig:4} also illustrates the performance of two antenna patterns (Patch and Yagi) mounted at the SC. 
It can be observed that SCs equipped with a more directive antenna for access (i.e. Yagi) provide higher performance than those equipped with the isotropic antenna (i.e. Patch). The performance enhancement is caused by two complementary effects: \emph{i)} the signal improvements provided by the larger antenna gain of the directive Yagi, and \emph{ii)} the reduced interference created towards neighbouring UEs served by other SCs. 

%The latter is a direct consequence of the reduced antenna gain towards the angular locations located far from the main beam in the directive antenna pattern. 
%Overall, the results of Fig. \ref{fig:4} allow us to conclude that having a dense network of SCs is beneficial for the s-BH system when an ad-hoc deployment is in place.

\subsection{Impact of the resource allocation} \label{subsec: RA results}

\begin{figure}[!t]
	\centering
	\subfloat[$5$-th percentile]{\includegraphics[width=0.9\columnwidth]{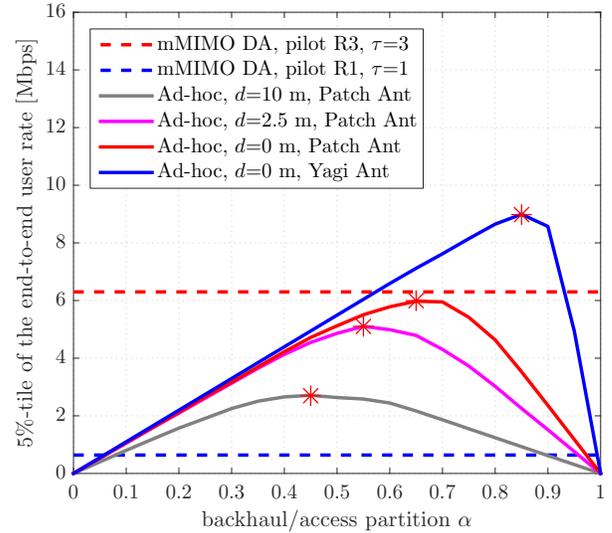}\label{fig:5a}}\hfill
	\subfloat[$50$-th percentile]{\includegraphics[width=0.9\columnwidth]{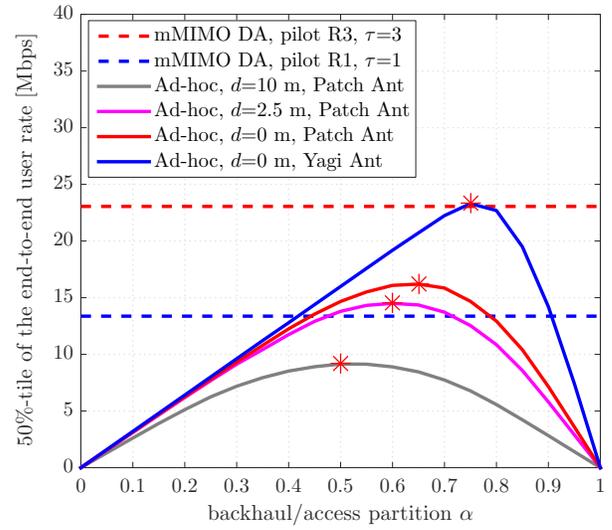}\label{fig:5b}}	 
	\caption{(a) $5$-th, and (b) $50$-th percentile of the UE rates as a function of the partition $\alpha$ between backhaul and access time-slots.}
	%\vspace*{-0.15cm}
	\label{fig:5} 
\end{figure} 

In Fig. \ref{fig:5}, we vary $\alpha$ in the range $0 \leq \alpha \leq 1$ to analyse the behaviour of UEs rate at the $5$-th and $50$-th percentiles of the CDF. The configurations $\alpha=0$ and $\alpha=1$ entail that all the time-slots are allocated to the access and the backhaul, respectively. Therefore, the UE rate for these two values is equal to 0, since no resources are left for the other link. Fig. \ref{fig:5} brings the following insights:

\begin{itemize}
	\item The s-BH links are generally the bottleneck of the two hop connection when the SCs are densely deployed near the UEs. 
	In such a case, in order to take advantage of the very high SINR experienced by the UEs on the access link, most of the resources must be allocated to the backhaul. For instance, with $d=0$ and Yagi antennas at the SCs, $\alpha^*$, i.e., the value of $\alpha$ that maximizes the UE rate, is about 0.85 when looking at the $5$-th percentile curve.
	
	\item By comparing the results between Fig. \ref{fig:5a} and Fig. \ref{fig:5b}, it is important to notice that the optimal $\alpha$ changes from 0.85 to 0.75. This deviation suggests that picking a non-optimal $\alpha$ can lead to a significant reduction of the end-to-end UE rates. In fact, assuming that the network uses $\alpha = 0.85$, which is the optimal value for cell-edge UEs ($5$-th percentile of the CDF), the median UEs ($50$-th percentile of the CDF) can achieve an end-to-end rate of 19.5 Mbps, which represents a $16\%$ reduction with respect to the maximum end-to-end rate achievable of 23.3 Mbps.
	
	\item In Figs. \ref{fig:5a} and \ref{fig:5b}, we show with dash lines the results of the mMIMO DA setup, which is considered as the baseline for the network performance. 
	The results show that a properly designed s-BH system can improve the performance of the cell-edge UEs, but this is not the case for the median-UEs. 
	A more detailed comparison is further developed in the next section.
\end{itemize}

%\item Picking a non-optimal $\alpha$ leads to degraded network performance. Indeed, it can be observed that $\alpha^{\ast}$ changes from 0.85 to 0.75 between Fig. \ref{fig:5a} and Fig. \ref{fig:5b}. This suggests that the median UEs ($50$-th percentile of the CDF) can experience a $20\%$ reduction of the end-to-end rate with respect their maximum rate when the network uses $\alpha = 0.85$, which is the optimal value for cell-edge UEs ($5$-th percentile of the CDF). 
%\item In Fig. \ref{fig:5} are also reported the results of the mMIMO DA setup, considered here as a baseline, which show that only a properly optimized s-BH system can outperform DA at the cell edge, while guaranteeing performance comparable to the DA setup for the median UEs. A more detailed comparison is further developed in the next section.

\subsection{Comparison between DA and s-BH systems}

\begin{figure}[!t]
	\centering
	\includegraphics[width=0.9\columnwidth]{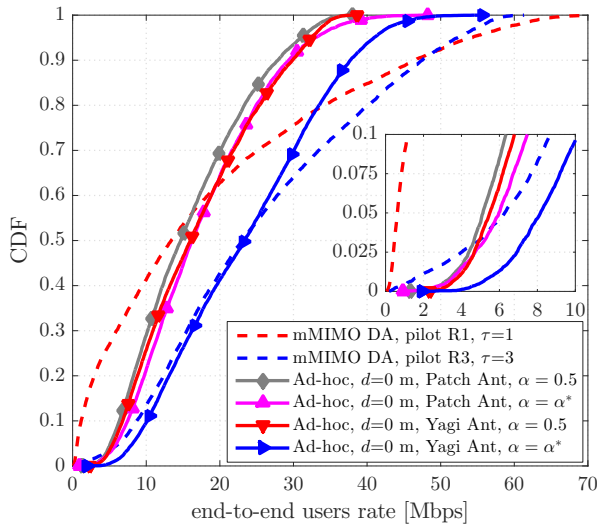}
	\caption{Two types of curves are represented: ($i$) mMIMO DA with pilot reuse schemes 1 and 3; ($ii$) ad-hoc deployment of 16 SCs per sector for $\alpha=0.5$ and $\alpha=\alpha^{\ast}$, at which the $50$-th percentile of the UE rate is maximized (as shown in Fig. \ref{fig:5b}).}
	\label{fig:6}
\end{figure}

In this section, we further compare the mMIMO s-BH and the mMIMO DA architectures to identify in which conditions the network results in a better performance in terms of UE rates. For the mMIMO DA architecture, $K_{i}=16$ UEs are trained and served per time-slot $T$. From Fig. \ref{fig:6}, we identify two different regions:
\begin{itemize}
	
	\item At the bottom of the CDF, i.e. below the $50$-th percentile, the mMIMO s-BH architecture with $d=0$ and Yagi antennas provides better performance when compared to the mMIMO DA architecture with pilot reuse 1.
	This is because pilot contamination severely degrades the rate of UEs at the cell edge in the mMIMO DA setup with pilot reuse 1. In this region, s-BH architecture works better because: 1) access links benefit from the UE-to-SC proximity, which reduces the path loss and improve the LOS propagation condition, and 2) backhaul links benefit from the absence of pilot contamination, and the higher height of the SC compared to the UE. The latter leads to an improved path-loss and LOS conditions with respect to those modelled for the macro to UE link \cite{3gpp.36.814}.
	%as reflected in the modified models for LOS and path-loss compared to those for macro to UE link \cite{3gpp.36.814} and from the absence of pilot contamination. 
	The gain achieved by mMIMO s-BH decreases when mMIMO DA with pilot reuse 3 is considered, mainly because of the reduced pilot contamination effect in the latter. 
	However, we still observe some considerable gain of about $30\%$ with $d=0$ and Yagi antennas when looking at the 5-th percentile of the UE rate.
	
	\item At the top of the CDF, i.e. over the $50$-th percentile, the mMIMO DA architecture exceeds the performance of s-BH mMIMO. This is because, with s-BH, the end-to-end rates are conditioned by both inter-cell interference and limited backhaul capacity, which combined set a limit to the maximum rate of the two hop connection. 
\end{itemize}
The results of Fig. \ref{fig:6} confirm that the deployment of s-BH architectures can be effective for serving UEs located at the cell edge and motivate the use of mMIMO DA for serving median UEs.

\section{Conclusion} \label{Sec:5}
In this paper, we analyzed the performance of the mMIMO based s-BH architecture below 6 GHz frequencies. We adopted a system-level simulation approach to investigate the UE rate performance for different SC deployments, and to analyze the effect of the variation of the backhaul/access partition. 
We showed that end-to-end performance greatly benefits from an ad-hoc SC deployment with one SC per UE, and studied the optimal backhaul/access partition, which maximizes the end-to-end rates, for the cell-edge UEs and for the median UEs.
Additionally, we also study the effect of that antenna directivity adopted at the SCs, which leads to important gains in a dense deployment of SCs. 
Finally, we compared the s-BH architecture against a mMIMO DA baseline. By properly optimizing the SC deployment and antenna directivity, s-BH outperforms DA at the cell edge in ultra-dense deployments, in particular when pilot reuse 1 is used in the latter. On the other hand, DA outperforms s-BH when looking at the median UEs. 

%Future works will address the effects produced by a variation in the SC deployment height, which may lead to more favourable propagation conditions in the backhauling link, thus potentially further boosting the performance of the mMIMO s-BH architecture.

\bibliographystyle{IEEEtran}
\bibliography{Strings_Gio,mycollection}

\end{document}